\documentclass[manuscript,screen]{acmart}
\AtBeginDocument{%
  \providecommand\BibTeX{{%
    \normalfont B\kern-0.5em{\scshape i\kern-0.25em b}\kern-0.8em\TeX}}}

\setcopyright{none}
\copyrightyear{2024} 
\acmYear{2024} 
\acmDOI{} 

\acmConference[LART@CHI '24]{LLMs as Research Tools: Applications and Evaluations in HCI Data Work, a workshop at CHI 2024}{May 12th, 2024}{Honolulu, HI}

\begin{document}

\title{Apprentices to Research Assistants: Advancing Research with Large Language Models}

\author{Mohammad Namvarpour}
\email{matt.namvarpour@drexel.edu}
\affiliation{
 \institution{Drexel University}
 \city{Philadelphia}
 \state{Pennsylvania}
 \country{USA}}
\author{Afsaneh Razi}
\email{afsaneh.razi@drexel.edu}
\affiliation{
 \institution{Drexel University}
 \city{Philadelphia}
 \state{Pennsylvania}
 \country{USA}}

\renewcommand{\shortauthors}{Namvarpour and Razi}

\begin{abstract}
Large Language Models (LLMs) have emerged as powerful tools in various research domains. This article examines their potential through a literature review and firsthand experimentation. While LLMs offer benefits like cost-effectiveness and efficiency, challenges such as prompt tuning, biases, and subjectivity must be addressed. The study presents insights from experiments utilizing LLMs for qualitative analysis, highlighting successes and limitations. Additionally, it discusses strategies for mitigating challenges, such as prompt optimization techniques and leveraging human expertise. This study aligns with the \textit{`LLMs as Research Tools'} workshop's focus on integrating LLMs into HCI data work critically and ethically. By addressing both opportunities and challenges, our work contributes to the ongoing dialogue on their responsible application in research.
\end{abstract}

\ccsdesc[500]{Computing methodologies~Machine learning}
\ccsdesc[300]{Human-centered computing~Human computer interaction (HCI)}
\ccsdesc[300]{Information systems~Information extraction}
\ccsdesc[100]{Social and professional topics~Codes of ethics}
\ccsdesc[500]{Applied computing~Document management and text processing}

\keywords{Large Language Models (LLMs), Natural Language Processing (NLP), Research Tools, Artificial Intelligence (AI)}


\received{22 February 2024}
\received[accepted]{22 March 2024}

\maketitle

\section{Our new lab members?}



Although the surprising abilities of Large Language Models (LLMs) only became public knowledge after the introduction of ChatGPT and the ensuing excitement, for those of us who have been following research around LLMs, it was an emerging capability that we could foresee since the introduction of GPT-2 \cite{Radford2019LanguageMA}. However, the LLMs were not reliable enough to wear the white jacket and assist in our research.

Ever since ChatGPT was introduced \cite{OpenAI_2022}, and subsequently tested in every imaginable scenario, from passing SAT exams \cite{webbEmergentAnalogicalReasoning2023} to boost productivity in white-collar jobs \cite{noyExperimentalEvidenceProductivity2023}, the research community has also become bolder. This might have been due to the push from PhD research assistants, who tried to convince their hesistant advisors that the time had come for them to offload some of the less critical and more mundane aspects of their job (coding, relevancy coding, summarization, etc.) to LLMs. After all, if lawyers \cite{cnbc2023} and editors of famous magazines \cite{pbs2023} can try that, why can't we?

So far, utilization of LLMs in research have been the topic of several studies. In a comprehensive evaluation conducted by Bang et al. \cite{bangMultitaskMultilingualMultimodal2023}, ChatGPT was put through a multitask, multilingual, multimodal evaluation. The study found that ChatGPT demonstrated proficiency in several areas including misinformation detection and sentiment analysis. However, the same study also highlighted some limitations of ChatGPT, including struggling with inductive, spatial, mathematical, and non-textual semantic reasoning. Thus, while ChatGPT showed promise in certain areas, it also revealed areas that require further improvement.
ChatGPT also proved valuable in annotating implicit hate speech \cite{huangChatGPTBetterHuman2023} and achieved acceptable accuracy rates in categorizing statements as true or false for fact-checking \cite{hoesLeveragingChatGPTEfficient2023}. Similarly, stance detection tasks \cite{zhangHowWouldStance2023} saw ChatGPT producing top-tier results, revolutionizing the field. However, its non-deterministic nature \cite{reissTestingReliabilityChatGPT2023} led to inconsistent text annotation and classification outputs, emphasizing the need for validation and caution in unsupervised applications. Concerns were also raised about the potential for data leakage affecting model evaluations \cite{aiyappaCanWeTrust2023}. Despite these concerns, Gilardi et al. \cite{gilardiChatGPTOutperformsCrowdWorkers2023} found that ChatGPT excelled in text-annotation tasks, surpassing crowd-workers in terms of accuracy and cost-effectiveness.

In a study by Ziems et al. \cite{ziemsCanLargeLanguage2023}, the usage of LLMs for textual analysis in various computational social science tasks (e.g., Political Science, Psychology, Sociology) was explored. Results showed LLMs excel in classifying shorter texts with high accuracy and agreement with human annotators, while occasionally outperforming humans in text generation. However, the paper suggested that LLMs should only be used as supplementary annotators and in an augmentative capacity. Researchers were cautioned against solely relying on LLMs for their data analysis.

While the insights from existing studies are valuable in understanding the utility of LLMs in research, offering a firsthand account can provide a more engaging and practical narrative for readers of this article. Therefore, in the following section, we explain our experience with LLMs in a recent project, detailing our observations and approaches.

\section{Our experiments with Large Language Models}

In a recent research project, we decided to try the new Generative AI capabilities as a part of our qualitative study pipeline to help reduce the human labor needed for the project. The corpus on which we wanted to work was comprised of about 35 thousand short spans of text, covering a wide range of topics. We only wanted to study a thin slice of data in which authors mentioned their experience of unwanted online sexual behavior and harassment; a task that was similar to looking for a needle in a haystack.

We initially adopted a basic keyword search for text related to online sexual harassment, refining our keyword list through iterative trials to optimize text retrieval. While this method might be acceptable for searching among more `formal' types of text such as papers and news articles, it certainly was not appropriate for our informal, user-generated content. In fact, our goal in identifying relevant keywords was to build a small corpus of text related to online sexual harassment. We compiled a corpus of 500 text spans. To achieve this, we randomly picked 250 text spans from a subset of our data containing online sexual harassment-related keywords. The remaining 250 spans were selected randomly from the entire dataset to ensure a representative sample. Subsequently, two human annotators labeled each text span as either relevant or non-relevant to our topic of online sexual harassment. They achieved near-perfect agreement with a Cohen's kappa score of 0.994. We concluded that the task is straightforward and can be automated. The resulting corpus could be used to train a model, thereby automating the process of filtering out relevant text spans.

We initially chose a logistic regression classifier based on BERT \cite{devlinBERTPretrainingDeep2019b} to identify relevant text spans. However, due to our small dataset mainly comprising spans with specific keywords related to online sexual harassment, we were concerned the model might overfit and merely recognize spans containing those keywords. Consequently, we adopted an active learning strategy, a machine learning subset where the algorithm queries a human for labeling uncertain data points, aiming to refine its decision-making without large datasets \cite{cohnImprovingGeneralizationActive1994}. Despite the engaging learning experience, the approach didn't significantly improve our model. After processing 35,000 text spans, the model showed high precision but very low recall, accurately identifying few relevant instances and marking most data as irrelevant.

After our first experiment failed, we decided to utilize the new tools in our toolkit, the OpenAI’s GPT models. To ensure the quality of GPT’s annotations, we first wanted to classify our dataset of 500 text spans and compare it with human annotations. To achieve this, we engineered the following prompt for the Language Models:

\textit{Given the text below, determine if it contains a complaint about online sexual harassment. For the purpose of this task, ‘Online Sexual Harassment’ is defined as any unwanted or unwelcome sexual behavior on any digital platform using digital content (images, videos, posts, messages, pages), which makes a person feel offended, humiliated or intimidated. If the review contains a complaint about such behavior, such as unsolicited flirting or inappropriate texts that are sexual in nature or otherwise harassing, output 1. Otherwise, output 0.: [here goes the review]}

We tested both GPT3.5-turbo \cite{openaiGPT35turbo2023} and GPT4 \cite{openaiGPT42023} through OpenAI’s API and measured their performance against human annotations using the Fleiss Kappa Score. GPT3.5-turbo scored 0.799, while GPT4 scored 0.837. Both models showed a strong agreement with human annotations, but GPT4 tended to follow instructions more accurately. For example, while it was mentioned in the prompt that we expected the output to be a single digit, GPT3.5-turbo would sometimes return an explanation, or answers such as ‘Yes’ and ‘No’. GPT4, on the other hand, almost always followed the requested format. However, GPT4 was more expensive to use. Since GPT3.5-turbo's results were still satisfactory, we opted for it due to its cost-efficiency. We input our dataset of 35 thousand text spans into GPT3.5-turbo, which identified 1040 text spans related to online sexual harassment. We then conducted a thematic analysis on this subset of data, performed manually by humans, which is not discussed further in this article.

Upon manual review of the classified text spans, we observed that the language model successfully identified text spans that would pose significant challenges for other machine learning models. Given the sensitivity of our study topic, many instances avoided using exact phrases, opting for substitutions like `seggs' for `sex' and `faq' for `fuck', making it difficult for simpler models like our initial BERT-based classifier to comprehend. While impressed by the natural language understanding capabilities of LLMs, we found that approximately a quarter of the classifications were false positives. This means that although GPT3.5-turbo identified them as relevant to online sexual harassment, they were not. However, in most cases, these false positives still discussed issues related to online sexual behavior, albeit not as complaints and thus not categorized as `harassment'.

We applied LLMs in the relevancy coding phase, condensing 35,000 text spans into 1040. Despite around a quarter not directly relevant, this smaller dataset sufficed for our qualitative analysis and theme extraction. Our experiment with LLMs was successful and replicable. Despite some inaccuracies, they didn't hinder our study. In the next section, we'll discuss the opportunities and challenges of using LLMs in research.

\section{Research Assistant LLMs: Opportunities and Challenges}

Up to this point, we have presented existing research on the utilization of LLMs in the first section, followed by our personal encounters with LLMs in the second section. In this concluding segment of the article, we summarize the opportunities and challenges associated with employing large language models as a tool in research.

Considering how crucial data is in many research endeavors, LLMs have shown great promise in aiding research efforts. They've often performed better than crowdsourced workers \cite{gilardiChatGPTOutperformsCrowdWorkers2023}, who are commonly used for tasks like data analysis. With LLM technology advancing steadily and without major setbacks so far, it seems likely that LLMs will start taking over many tasks, especially simpler ones, from human workers. 
This presents an opportunity for researchers to conduct larger-scale studies with better quality and lower cost compared to just a few years ago.

However, there are concerns that need to be carefully considered before fully embracing LLMs in research. One of the most challenging aspects of working with LLMs is prompt tuning. Unfortunately, prompts play a crucial role in determining how well the model performs. By utilizing techniques like chain-of-thoughts \cite{weiChainofThoughtPromptingElicits2022} and retrieval augmented generation (RAG) \cite{lewisRetrievalaugmentedGenerationKnowledgeintensive2020}, along with selecting better prompts, the performance of a specific model can be significantly enhanced without altering the model itself. This highlights that while the model may inherently possess the ability to excel in a particular task, this potential cannot be fully realized without employing specific methods and prompts. Such reliance on external factors introduces a high level of instability and non-reproducibility, which can hinder the integrity of research outcomes.

In addressing this issue, Khattab et al. \cite{khattabDSPyCompilingDeclarative2023} have introduced the DSPy library, offering researchers a more systematic approach to utilizing LLMs. This library includes an automatic teleprompter feature, which can generate prompts for researchers. While their approach does not negate the significance of prompts and techniques in enhancing LLM performance, it assists researchers, particularly those less experienced in prompt optimization, in obtaining favorable results from their models. Additionally, their work demonstrates that with an effective prompt, even simpler, open-source models can yield satisfactory outcomes.

Another significant challenge lies in addressing the biases inherent in LLMs, which can be particularly problematic when these models are applied to computational social science tasks. A concerning discovery highlighted by Ziems et al. \cite{ziemsCanLargeLanguage2023} underscores this issue. They conducted experiments using various LLMs, including both open-source models and OpenAI's GPT models. In some instances, they observed unanimous errors, where all models would agree on a false answer. This phenomenon, termed unanimous error, suggests that due to the similarity of datasets used to train large models, they may exhibit similar biases and errors that are not automatically detected, as they are shared among other LLMs as well. Unfortunately, as demonstrated by Glaese et al. \cite{glaeseImprovingAlignmentDialogue2022}, even human feedback alone does not entirely eliminate biases either and may sometimes exacerbate them.

Another critical aspect to consider regarding the utilization of LLMs in research is the nature of the research itself. Some research endeavors are inherently more subjective and interpretive. In these cases, the beliefs and lived experiences of researchers significantly influence the outcome of their research and should be acknowledged when publishing results. With the widespread adoption of LLMs, the scientific community risks losing the diverse perspectives of different researchers investigating the same problem. By relying on LLMs as data annotators and analyzers, researchers' perspectives are obscured, and their unique insights may not be fully reflected in their experiments and results. Therefore, the use of LLMs in such fields should be approached with heightened caution.

\section{Conclusion}
Given the rising opportunities and challenges brought about by LLMs in research, it's crucial to carefully examine their methodological and ethical implications in the HCI community. We believe that LLMs can significantly improve research efficiency and data analysis. However, we also believe we need to consider their biases and the importance of human oversight carefully. This viewpoint not only supports the objectives of the \textit{`LLMs as Research Tools'} workshop, which aims to ensure methodological validity and address ethical concerns in HCI data work but also contributes to the broader discussion on responsibly integrating LLMs into research practices.
By sharing our experiences and insights, we aim to help establish community standards and enhance understanding of LLM applications in HCI research. Our work serves as a case study, illustrating the potential benefits and challenges of integrating LLMs, thus adding to the collective knowledge base on how to navigate LLM use in research settings. Through this contribution, we hope to promote a balanced and critical approach to using LLMs as research tools, furthering the workshop's goal of outlining current approaches and addressing challenges in this rapidly evolving field.

\bibliographystyle{ACM-Reference-Format}
\bibliography{refs}










\end{document}